# Probing Millikelvin Temperature Sensitivity in Chiral Nanoparticles via Optical Forces

Seongmin Im[1], Wei Hong[1,2,3,4], Gayatri Chandran[1,5], Xing Wang[1,2,3,4,6], and Yang Zhao[1,2,4,5,7,8,*]

[1]Nick Holonyak Micro and Nanotechnology Laboratory, University of Illinois Urbana-Champaign, Urbana, IL 61801, USA

[2] Department of Bioengineering, University of Illinois at Urbana-Champaign, Urbana, IL 61801, USA

[3]Department of Chemistry, University of Illinois at Urbana-Champaign, Urbana, IL 61801, USA.

[4]Carl R. Woese Institute of Genomic Biology, University of Illinois Urbana-Champaign, Urbana, IL 61801, USA

[5]Department of Electrical and Computer Engineering, University of Illinois Urbana-Champaign, Urbana, IL 61801, USA

[6]Cancer Center at Illinois, University of Illinois at Urbana-Champaign, Urbana, IL 61801, USA.

[7]Beckman Institute, University of Illinois Urbana-Champaign, Urbana, IL 61801, USA

[8]Translational Sciences, Carle Illinois College of Medicine, University of Illinois Urbana-Champaign, Urbana, IL 61801 USA

*Corresponding author: yzhaoui@illinois.edu

**Abstract:** With increasing interest in utilizing nanostructures as nanoscale heat sources, the ability to precisely measure photothermal effects at the nanoscale has become increasingly significant. Techniques based on fluorescence or Raman signals often suffer from challenges in accurate calibration, far-field imaging methods are limited by diffraction-limited spatial resolution, and electron microscopy requires vacuum conditions, restricting in situ applicability. In contrast, tip-based measurement techniques offer sub-diffraction spatial resolution under ambient conditions, making them well-suited for nanoscale photothermal mapping. In this study, we employ tip-based optical force nanoscopy combined with phase-informed decomposition to investigate the origin of the photothermal force, enable nanoscale mapping, and evaluate temperature sensitivity. Our system achieves a temperature sensitivity of approximately 0.1 K without necessitating an additional temperature-sensitive layer. We anticipate that our approach has the potential to serve as a versatile platform for investigating localized thermal effects in fields such as semiconductors, nanophotonics, and photocatalysis.

**Keywords:** Optical force nanoscopy, photothermal force, thermal expansion, chiral nanoparticle

# Introduction

Temperature is a physical quantity that describes the motion of atoms or molecules within a material and indicates its internal energy. However, accurate temperature measurement remains a significant challenge due to the rapid dissipation of thermal energy and the difficulty in maintaining thermal equilibrium. Nanophotonics has recently emerged as a versatile platform for sensing[1, 2], imaging[3], photocatalysis[4], energy harvesting[5], and biomedical applications[6, 7]. Beyond its ability to concentrate and enhance light-matter interactions, nanophotonics also enables the conversion of optical energy into various forms. Traditionally, the photothermal conversion of light into heat has been considered a parasitic loss. However, the growing interest in employing nanoparticles as localized heat sources[8, 9] has highlighted the importance of nanoscale temperature visualization, particularly in emerging applications such as photoacoustic imaging[10], photothermal catalysis[11, 12], and photothermal therapy[13].

To achieve microscopic temperature mapping, various fluorescence-based thermometry techniques have been developed, which rely on temperature-dependent changes in fluorescence intensity[14], lifetime[15], and anisotropy[16]. Despite their potential, these methods face inherent limitations. Fluorescence has a low dynamic range in temperature due to the loss of signal characteristics at high temperatures and is also affected by various extrinsic factors, including pH, molecular concentration, surface chemistry, and energy transfer processes[17], complicating accurate temperature calibration. Raman thermometry faces comparable constraints[18]. Alternative approaches include quantitative phase microscopy[19], which visualizes temperature through temperature-induced changes in the refractive index, and photothermal microscopy[20], which employs thermal lensing effects along with pump-probe measurements. While these methods enable cellular-scale thermal imaging and, in the case of photothermal microscopy, can be combined with infrared spectroscopy for chemical contrast, they are inherently limited by diffraction due to their far-field imaging nature. Electron energy loss spectroscopy overcomes the diffraction limit and enables temperature mapping with nanometer-scale spatial resolution[21]; however, its applicability is constrained by the need for electron microscopy under high-vacuum conditions.

In contrast, tip-based thermometry techniques offer superior spatial resolution determined by the tip size, allowing access to sub-diffraction-limited thermal features. Atomic force microscopy-infrared and photo-induced force microscopy, which combine atomic force microscopy with infrared spectroscopy, enable simultaneous acquisition of thermal expansion and chemical information at the nanoscale[22, 23, 24, 25, 26]. Apertureless near-field scanning optical microscopy revealed that photothermal expansion of PMMA was the predominant effect, surpassing the contribution of optical gradient forces in the infrared region[27]. In addition, aperture-type near-field scanning optical microscopy allows direct observation of the nonlinear features in light absorption and near-field enhancement[28]. Recently, decoupled optical force nanoscopy has been introduced, employing phase-informed decomposition to distinguish the optical gradient and photothermal force[29]. This technique has demonstrated the potential for investigating ultrafast photothermal response with high spatial resolution.

Here, we present a methodology for temperature quantification based on the correlation between photothermal force and thermally induced expansion, as measured by optical force nanoscopy. We systematically investigate the underlying origin of the photothermal force and derive the theoretical temperature sensitivity limit of the technique. Leveraging this framework, we further demonstrate polarization-dependent thermal responses in chiral gold nanoparticles under circularly polarized light, highlighting the capability of the approach for nanoscale thermometric analysis with sub-diffraction spatial resolution.

# Results and discussion

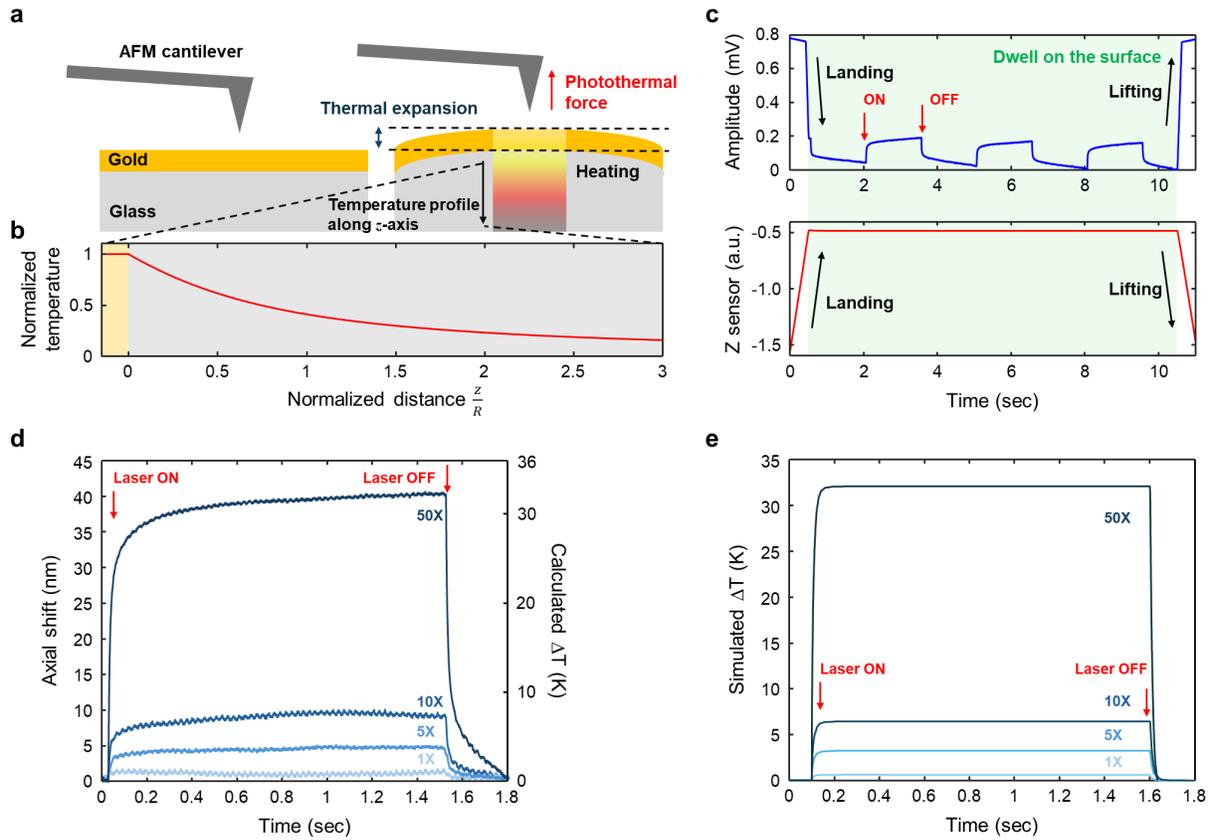

**Figure 1. Origin of photothermal force. a** Schematic illustration of the photothermal force, showing thermal expansion upon optical heating of a gold film and the underlying glass substrate. Heat dissipates through the substrate, and a scanning probe measures the overall thermal expansion of the film. **b** Analytical solution of the temperature profile along the *z*-axis under uniform illumination. R denotes the radius of the illuminated field of view. **c** Amplitude of the cantilever deflection and z sensor signal as functions of time during a cycle in which the probe engages, dwells, and withdraws. During the dwell phase, the laser is toggled on and off to allow measurement of the photothermal force. **d** Measured temporal response of the axial shift of the gold film under various incident laser powers. "1X" corresponds to an incident power of $1.5 \times 10^5$ W/m². The numbers indicate power

ratios relative to this baseline. **e** Simulated temporal response of the gold film under varying incident laser powers.

## Thermal expansion and photothermal force

It might seem unusual that the thermal expansion of a metal film just a few tens of nanometers thick can result in a measurable displacement of several tens of nanometers. However, as the temperature of the glass substrate onto which the metal film is deposited increases, the thermal expansion of the glass substrate itself can also reach several tens of nanometers. The typical thickness of a glass substrate is around 150 μm. The thermal expansion coefficient of a BK7 glass substrate is 7.1 × $10^{-6}$/K, meaning that a uniform increase of 1 °C in temperature leads to an expansion of approximately 1 nm.

Figure 1a shows a schematic illustration of photothermal expansion and the resulting photothermal force of a gold thin film deposited on a glass substrate. When light is absorbed by the gold film, it generates heat, leading to the thermal expansion of both the gold film and the glass substrate. The AFM cantilever maintains a constant distance between the tip and the sample, where the deflection signal reflects the photothermal force. Figure 1b illustrates the temperature profile along the z-axis under uniform illumination, which is obtained from Equation (1)[30] in cylindrical coordinates:

$$T(z) = \frac{q_0}{4\pi\kappa} \int_0^R \int_0^{2\pi} \frac{1}{\rho^2 + z^2} \rho d\theta d\rho = \frac{q_0}{2\kappa}(\sqrt{R^2 + z^2} - z) \quad (1)$$

where $q_0$ and $\kappa$ represent the heat source and thermal conductivity, respectively. R is the radius of the illuminated area. The resulting thermal expansion can be calculated using Equation (2)[28]:

$$L = L_{gold}\left(1 + 3\alpha_{gold}\Delta T(z)\right) + \int_{-L_{gold}-L_{BK7}}^{-L_{gold}} \left(1 + 3\alpha_{glass}(T_{\max} - T_{\min})\left(\sqrt{1 + \left(\frac{z}{R}\right)^2} - \frac{|z|}{R}\right)\right) dz \quad (2)$$

where $L_{gold}$ and $L_{BK7}$ represent the thickness of the gold film and the glass substrate, respectively. α denotes the thermal expansion coefficient of the materials. The thermal expansion coefficients used in the analysis are 14 × 10$^{-6}$ 1/K for gold and 7.1 × 10$^{-6}$ 1/K for glass. Since the illuminated area covers only a small portion of the total sample surface, the volumetric expansion is approximated as linear expansion.

To verify the thermally induced axial shift of the AFM cantilever, we positioned the cantilever onto the sample surface and cycled the laser on and off at intervals of 1.5 seconds. The experiments were performed on a 160 nm gold film with zero transmittance to prevent the heating beam from being detected by the AFM detector. Circularly polarized light at a wavelength of 637 nm was used as the heating beam. As illustrated in Figure 1c, the z sensor, which indicates the tip-sample distance, maintains a constant value while the tip dwells on the surface (green-shaded area). Meanwhile, the amplitude signal, reflecting the relative z position of the cantilever, fluctuates up and down depending on whether the laser light is on or off. Figure 1d shows the axial shift of a 160 nm thick gold film on a glass substrate under varying incident powers. The amplitude-to-nanometer conversion coefficient (Amp InvOLS) was measured to be 306.89 nm/V. The measured axial shift increased with increasing power. The lowest power, denoted as 1X, corresponds to an incident power of $1.5 \times 10^5$ W/m$^2$. At 1X power, the average axial shift was 1.19 nm with a standard deviation of 0.223 nm. Increasing the power to 5X resulted in a mean shift of 4.65±0.228 nm, while at 10X power, the shift further increased to 10.38±0.547 nm. At the highest power level, 50X, the mean axial shift reached 39.26±0.921 nm. In reported values, the number preceding the ± symbol represents the mean, while the number following it denotes the standard deviation. The standard deviation relative to the mean stays within approximately 2–5% for power levels above 5X. The axial shift of 39.26 nm corresponds to the temperature increment of 32 K, which is in excellent agreement with the simulation results shown in Figure 1e. Our system enables the detection of an axial shift as small as 1.19 nm at the lowest power, corresponding to a temperature elevation of 0.9 K. The temperature detection limit of the optical force nanoscopy system is closely related to the axial resolution of the AFM, which is approximately

0.05 nm. Taking into account the thermal expansion of the glass substrate and the gold thin film, this resolution corresponds to a temperature change of approximately 0.04 K. While this value represents the theoretical lower limit of the system's temperature sensitivity, such minute temperature variations are highly susceptible to environmental fluctuations and thermal noise. Therefore, these measurements should be interpreted with caution.

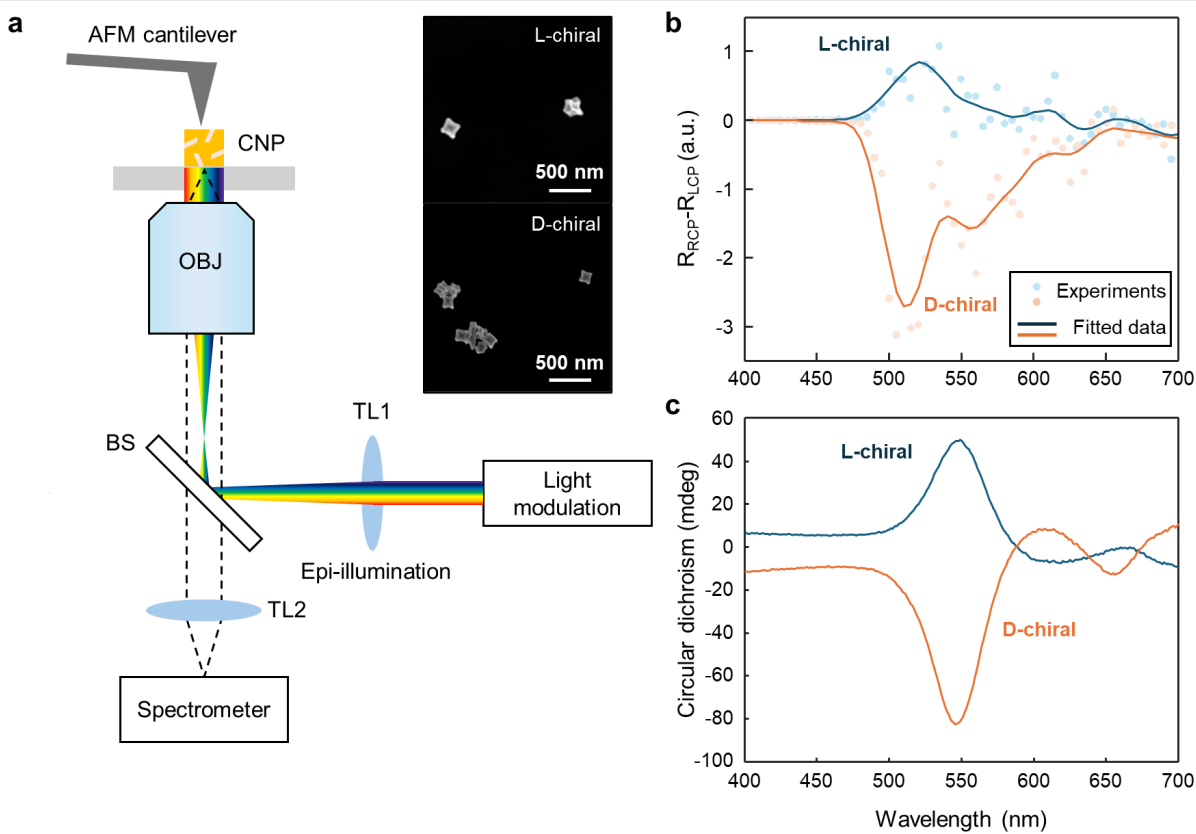

**Figure 2. Optical force nanoscopy setup and characterization of chiral gold nanoparticles. a** Schematic illustration of the optical setup and SEM images of the chiral gold nanoparticles (CNP: Chiral nanoparticles, OBJ: Objective lens, BS: Beam splitter, TL: Tube lens). **b** Reflectance difference spectra of the chiral gold nanoparticles measured on a glass substrate. **c** Circular dichroism spectra of the chiral gold nanoparticles in aqueous solution.

## Characterization of chiral nanoparticles

To determine the temperature sensitivity of a single particle, we employed chiral gold nanoparticles in conjunction with circularly polarized light. The handedness of the circular polarization affects the absorption properties of the chiral gold nanoparticles, resulting in varying temperature increases. By measuring these temperature differences, we further evaluated the temperature sensitivity of the optical force nanoscopy system. In addition, a key advantage of using circularly polarized light is its ability to generate a consistent optical response regardless of the orientation of the gold nanoparticles.

The chiral gold nanoparticles were synthesized using a seed-mediated growth method using rhombicuboctahedron nanoparticles[31, 32]. A scanning electron microscopy (SEM) image of the chiral gold nanoparticles is presented in Figure 2a. The nanoparticle concentration is $5.6 \times 10^9$ particles/mL. The particles were diluted to 10 % concentration using deionized water and drop-cast onto a glass substrate, followed by solvent evaporation under vacuum. Residuals were then removed by rinsing with deionized water.

Figure 2a also shows the optical measurement setup used to characterize the optical properties of the chiral nanoparticles. A supercontinuum laser was used to generate a white light source centered at 550 nm with a spectral bandwidth of 300 nm. The beam passes through a polarization modulation module consisting of a linear polarizer, a half-wave plate, and a quarter-wave plate before reaching the sample. The incident light is wide-field illumination with normal incidence. Upon interacting with chiral nanoparticles, the reflected light is collected and analyzed using a spectrometer. Spectra were recorded 10 times with an integration time of 100 ms at 2-nm intervals using 10 sampling boxes, and the average value was subsequently calculated. To eliminate the influence of the substrate, the spectrum of a bare glass substrate without nanoparticles was also measured and used as a background for subtraction. The measured data was smoothed using a moving average filter.

The differential reflectance spectra between left- and right-handed circularly polarized light are presented in Figure 2b. The L-chiral nanoparticles exhibit an overall positive differential reflectance with a peak at 520 nm, while the D-chiral counterparts show a negative differential reflectance with a dip at 510 nm. These spectral features are blue-shifted compared to the liquid-state circular dichroism spectra obtained using a commercial spectrometer (JASCO J-1500), as shown in Figure 2c. This discrepancy arises from differences in the surrounding environment: the measurements in Figure 2b were conducted at the glass–air interface, whereas those in Figure 2c were performed in aqueous solution using a commercial circular dichroism spectrometer. There is no measurable difference in circular dichroism between L-chiral and D-chiral gold nanoparticles at a wavelength of 640 nm (see Figure 2c). However, we observed a small difference in differential reflectance at the same wavelength, as shown in Figure 2b. Since such small signal differences are useful for evaluating sensitivity, we chose a wavelength of 637 nm to induce photothermal force.

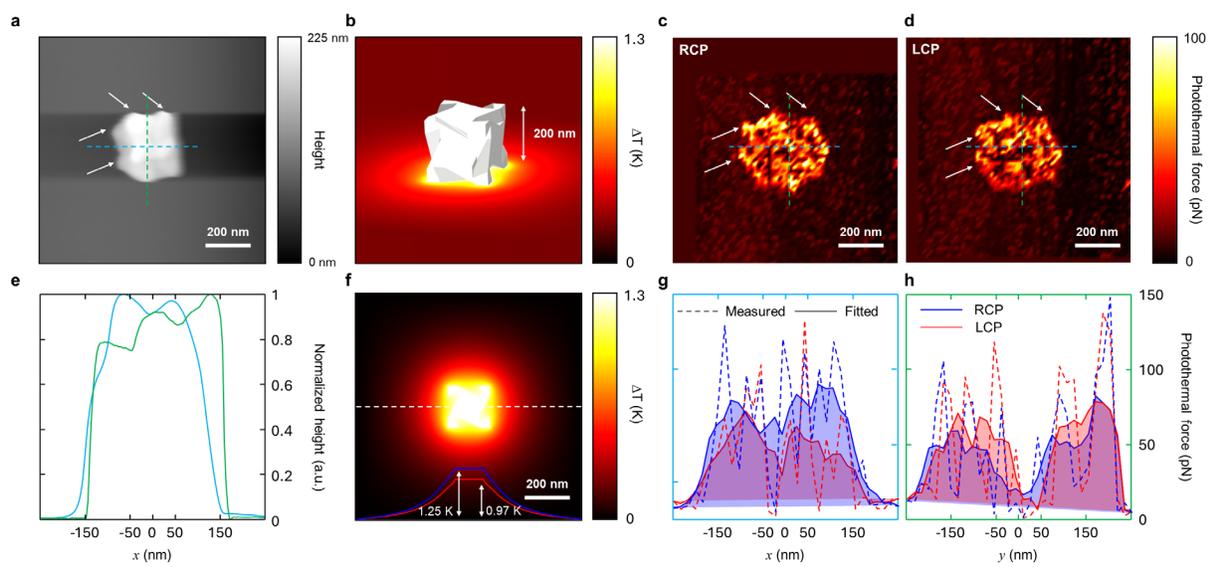

**Figure 3. Photothermal force mapping of an L-chiral gold nanoparticle. a** AFM topography of an L-chiral gold nanoparticle. **b** Simulated temperature distribution of the L-chiral gold nanoparticle under right-handed circularly polarized light (RCP). **c, d** Measured photothermal force maps of L-chiral gold nanoparticle under RCP, and left-handed circularly polarized light (LCP). **e** Height profile

of the L-chiral gold nanoparticle (*x*-axis: sky blue, *y*-axis: green). **f** Simulated temperature distribution at *z* = 200 nm and corresponding temperature profile along the *x*-axis (RCP: blue, LCP: red). **g, h** Measured photothermal force profiles along *x*- and *y*-axes. The scale bar represents 200 nm.

**Photothermal force of a chiral gold nanoparticle**

In the optical force nanoscopy system, the incident light was modulated using a function generator at a frequency $f_0$, which is identical to the AFM cantilever's resonant frequency. The AFM cantilever was driven at a frequency of 100 Hz above its resonant frequency to ensure operation in the attractive regime. This modulation enables the decomposition of the total optical force into three components: optical gradient force, photothermal force, and background force—typically attributed to photoacoustic effects—through phase-informed analysis[29]. Figure 3 presents the measured photothermal force from a L-chiral gold nanoparticle. Figure 3a shows the AFM topography of a single L-chiral gold nanoparticle. The arrows indicate the vertices of the chiral nanoparticle, which are also identifiable in the photothermal force map discussed later. The maximum measured height reaches up to 225 nm. Figure 3b presents a simulated temperature distribution of a 200 nm chiral cube that mimics the geometry of an L-chiral gold nanoparticle under right-handed circularly polarized light (RCP). The incident optical power was set to approximately $8 \times 10^6$ W/m$^2$, matching the experimental conditions. The thermal conductivities used in the simulation were 0.025 W/m·K for air, 317.98 W/m·K for gold, and 1.114 W/m·K for the BK7 glass substrate. A temperature rise of 1.25 K was observed under RCP illumination, whereas a comparatively lower increase of 0.97 K was observed under left-handed circularly polarized light (LCP).

Figure 3c, d show the photothermal force distribution of an L-chiral gold nanoparticle under RCP and LCP illumination, respectively. The maximum photothermal force was measured to be 142 pN for RCP, whereas 117 pN for LCP. The four vertices of the chiral nanoparticle, indicated by arrows in Figure 3a, are clearly identifiable in the photothermal force maps. Notably, the recessed areas of the chiral nanoparticle exhibit significantly lower photothermal force compared to the protruding features. Additionally, consistent with the simulation results, the photothermal force within the nanoparticle is substantially higher than in the surrounding regions. Figure 3e presents the topographic profiles along the *x*- and *y*-axes of the chiral nanoparticle. Along the *x*-axis, two protruding features are observed, while three distinct protrusions are noted along the *y*-axis. Although these structural features are not fully replicated in the simulation depicted in Figure 3f, Figures 3g and 3h clearly reveal corresponding photothermal force distributions—two protrusions along the *x*-axis and three along the *y*-axis. These patterns arise from variations in temperature distribution caused by air gaps within the nanoparticle structure, which are dependent on the shape of the particle. The profiles shown in Figures 3g and 3h were smoothed using a moving average filter. The average photothermal force over the field of view, excluding the nanoparticle region, was measured to be 13.3 pN under RCP and 15 pN under LCP. However, when focusing exclusively on the interior region of the chiral nanoparticle, the photothermal force significantly increases to 63.9 pN for RCP and 53.9 pN for LCP, indicating a pronounced polarization-dependent response.

This result suggests that heat dissipation is ultimately governed by the surrounding environment, and in the absence of a pronounced temperature gradient, the overall region experiences nearly uniform photothermal force and thermal expansion. The simulated temperature difference of 0.28 K corresponds to a photothermal force difference of approximately 11.7 pN, which was obtained from the difference in photothermal force between RCP and LCP. A photothermal force of 11.7 pN results in an observed axial shift of 0.37 nm, as determined by the back-calculation with spring constant of 2.5 nN/nm and frequency response of the cantilever. The resonant frequency of cantilever $f_0$ was 67.244 kHz, and the drive frequency of the cantilever $f_d$ was 67.344 kHz. In the analysis, the stiffening and softening of the cantilever due to the illumination were ignored. An axial shift of 0.37 nm corresponds to the thermal expansion of both the gold nanoparticle and the glass substrate, resulting from a 0.3 K temperature increase, consistent with the earlier analysis. This close agreement between the simulation and experiment confirms the consistency and validity of the results.

Importantly, the measured displacement surpasses the axial resolution of the AFM used in the system. With an AFM axial resolution of 0.05 nm, which corresponds to a temperature increment of 0.04 K based on thermal expansion calculations, we estimate that the minimum detectable optical force of the system is approximately 1.6 pN. However, we remain cautious about directly correlating absolute photothermal force values with precise temperature estimates. Although the temperature distribution within the gold nanoparticle is expected to remain uniform due to the high thermal conductivity of gold, the measured photothermal force also includes contributions from mechanical noise. This additional signal makes it challenging to assign a precise temperature to a given force value with high confidence. Therefore, determining temperature solely from individual photothermal force measurements may lead to inaccurate results. Instead, extracting temperature by comparing measured photothermal forces with calibrated reference data, as presented in this study, offers a more reliable and robust approach.

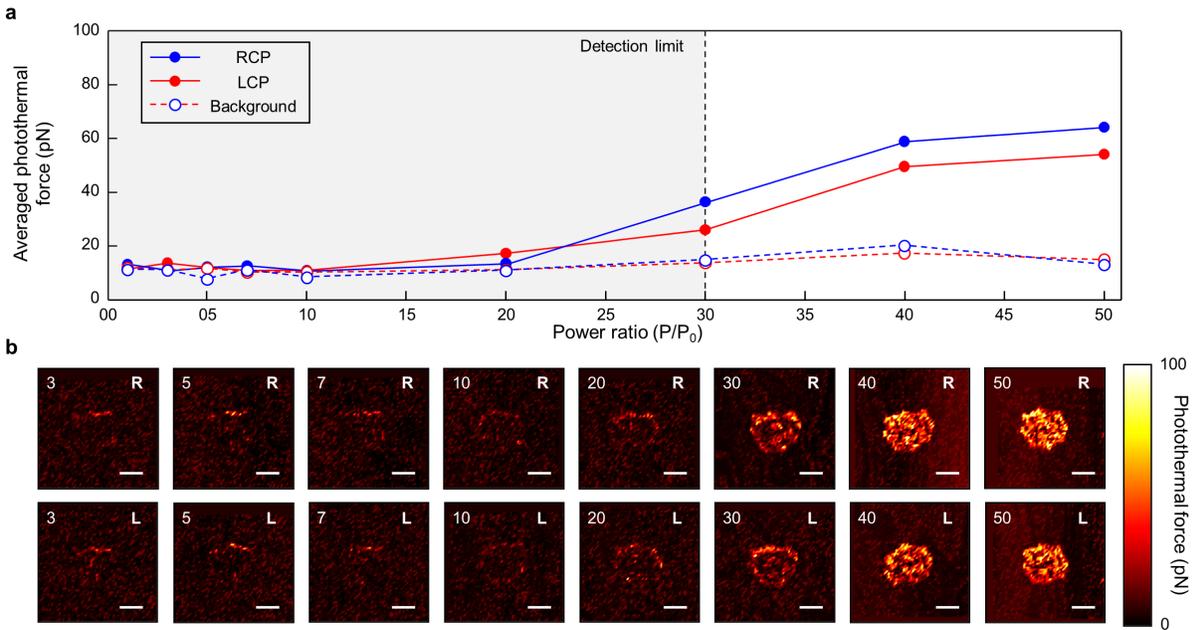

**Figure 4. Temperature sensitivity mapped on chiral gold nanoparticles. a** Average photothermal forces measured in the chiral gold nanoparticle region and the surrounding background as a function of incident optical power. The background signal was obtained from regions excluding the nanoparticle. To exclude areas of low photothermal force caused by internal air gaps within the nanoparticle, we calculated the average photothermal forces in two steps: first, a preliminary average was calculated over the nanoparticle region; then, a final average with the values that exceeded the preliminary average to exclude the air gaps. **b** Photothermal force maps under varying incident powers and circular polarizations. The top-left number in each panel indicates the power ratio, and the top-right R/L indicates the handedness of the circular polarization. The scale bar represents 200 nm.

**Temperature sensitivity of the optical force nanoscopy**

Figure 4 presents the average photothermal force and corresponding photothermal force maps within the chiral nanoparticle and across the entire field of view as functions of incident light power and polarization. The lowest power, used as a reference for calculating the power ratio, was $1.5 \times 10^5$ W/m$^2$. As previously discussed, Figure 4a confirms that the photothermal force within the L-chiral nanoparticle is higher under RCP. However, when the power ratio is below 30, it becomes difficult to distinguish the photothermal force originating from the nanoparticle from the background signal. This is due to an insufficient temperature rise, resulting in limited contrast in the photothermal force distribution. For example, at a power ratio of 20, the average photothermal force inside the nanoparticle is 13.3 pN for RCP and 17 pN for LCP, while the background levels are 11.1 pN for RCP and 11.2 pN for LCP. This leads to a modest difference of only 2.2 pN compared to the background level. In contrast, at a power ratio of 30, the photothermal force within the nanoparticle rises to 36.2 pN for RCP and 25.8 pN for LCP, with the background remaining at 13.6 pN for RCP and 13.8 pN for LCP. This results in a clear and reliable photothermal force difference of 7 pN between RCP and LCP.

This observation is further supported by the photothermal force maps illustrated in Figure 4b. At power ratios below 20, while the general shape of the nanoparticle is somewhat visible, it is more reasonable to attribute this to edge artifacts typically encountered in tip-based measurements rather than true photothermal force signals. In contrast, at power ratios of 30 and above, not only is the nanoparticle morphology clearly defined, but a uniformly high contrast photothermal force distribution is also observed within the particle.

Based on these findings, we define the temperature detection limit of the optical force nanoscopy system as the temperature difference between RCP and LCP, which is approximately 0.16 K. This value was obtained from simulation results using a power ratio of 30. From the measurements, the photothermal force difference of 7 pN corresponds to an axial shift of 0.22 nm. Based on thermal expansion calculation, this shift corresponds to a temperature increment of 0.18 K, corroborating the simulation results. Our analysis, illustrated in Figures 1, 3, and 4, demonstrates that the temperature detection limit of the optical force nanoscopy system is close to 0.1 K. However, it's important to note that this value may vary due to the inherent limitations of all tip-based measurements, specifically the dependence of data acquisition on the specific AFM tip used.

## Conclusion

In this study, we show that the origin of the photothermal force measured by optical force nanoscopy arises from the axial shift of the cantilever, which is primarily driven by the thermal expansion of both the nanoparticle and the substrate. The axial shift, obtained from experimentally measured photothermal force, can be converted into temperature through thermal expansion analysis. This inferred temperature shows excellent agreement with the predicted values obtained from photothermal simulations. In addition, we establish the temperature detection limit of the system based on the axial resolution of the AFM. The spatial resolution of our method is solely determined by the size of the tip, which readily surpasses the optical diffraction limit.

While our findings show that nanoscopic temperature mapping with a sensitivity on the order of 100 mK is achievable, several challenges remain to be addressed. Although optical force nanoscopy operates in the attractive regime, it remains sensitive to sample geometry, as evidenced by the edge artifacts observed in Figure 4b. Another limitation is the requirement for a reference point when measuring forces resulting from thermal expansion. If the entire field of view experiences a uniform temperature increase, axial shifts may occur, but without a relative baseline, these shifts would be indistinguishable. This constraint is not exclusive to our method; it is a common limitation among many indirect temperature detection techniques, most of which rely on comparing "on" and "off" states to establish a reference.

Our analysis sheds new light on tip-based thermal response measurements. In conventional nanoscale thermometric techniques, the contribution of substrate thermal expansion is often disregarded[33]. However, by employing phase-informed decomposition to selectively isolate the photothermal force, it becomes feasible to achieve temperature sensitivity on the order of 0.1 K without the need for an additional temperature-sensitive layer. This approach offers a distinct advantage by enabling high-resolution thermal measurements under ambient, non-vacuum conditions. Consequently, it holds significant promise for broad application in emerging fields where localized thermal effects at the nanoscale are critically important, such as semiconductor and nanophotonic device engineering, as well as photocatalysis.

# Materials and Methods

## Optical force measurements

The decoupled optical force nanoscopy system was implemented using an inverted optical microscope (IX81, Olympus) integrated with an atomic force microscope (MFP-3D-BIO, Oxford Instruments Asylum Research). A 637 nm diode laser (OBIS 637 nm, 140 mW, Coherent) was employed for measurements involving metallic thin films and photothermal force detection. The diode laser output was modulated in power, frequency, and duty cycle using a function generator (33250A 80 MHz Function/Arbitrary Waveform Generator, Agilent). The polarization state of the laser was controlled by the combination of a half-wave plate (AHWP10M-600, Thorlabs), and a quarter-wave plate (AQWP10M-580, Thorlabs). The beam was then focused onto the back focal plane of the objective lens (UAPON100XOTIRF, Olympus) via a tube lens, enabling wide-field illumination of the sample surface. An AC240 cantilever (4XC-NN, OPUS) was used for the AFM measurements. The deflection signal of the AFM cantilever was acquired using a lock-in amplifier (Signal Recovery 7280 DSP Lock-in Amplifier).

## Chiral nanoparticle synthesis

Au chiral nanoparticles were synthesized following a previously reported seed-mediated growth method using rhombicuboctahedron nanoparticles[31].

**Synthesis of 2 nm gold seeds.** In a typical synthesis, 32 mg of CTAC was dissolved in 975 µL of $H_2O$, followed by the addition of 25 µL of 10 mM $HAuCl_4$ solution. To this solution, 45 µL of freshly prepared ice-cold 20 mM $NaBH_4$ solution was rapidly injected under stirring, which was then kept still under 30 °C for 1 hour.

**Synthesis of Au RCO nanoparticles.** In a typical synthesis, two identical growth solutions (Vial A and Vial B) were prepared, each containing 320 mg of CTAC, 9.472 mL of $H_2O$, 250 µL of 10 mM $HAuCl_4$, and 30 µL of 1 mM KI. After mixing, 220 µL of 40 mM AA solution was added to each vial. After this, 55 uL of previously prepared 2 nm gold seed was added into Vial A under stirring. The solution changed to pink color within 10-15 seconds, and 55 μL of this mixture was then immediately injected into Vial B under stirring. Vial B was left undisturbed at 30 °C for 15 min to allow the growth of RCO nanoparticles. The RCO nanoparticles were cleaned twice using $H_2O$ by spinning at 6700 x $g$ for 2.5 min and dispersed in 1 mL of $H_2O$.

**Synthesis of Au chiral nanoparticles.** In a typical synthesis, 0.8 mL of 0.1 M CTAB and 100 μL of 10 mM $HAuCl_4$ were combined in a vial containing 3.6 mL of $H_2O$. To this vial, 475 μL of 0.1 M AA, 6 μL of 0.1 mM of L- or D- cysteine, and 3 μL of 1 mM $Cu(NO_3)_2$ were added sequentially. The mixture was gently mixed and incubated at 30 °C for 2 hours. The resulting nanoparticles were cleaned twice with $H_2O$ by spinning at 1000 x $g$ for 2 min and dispersed in 1 mL of $H_2O$.

### Ensemble circular dichroism measurement in aqueous solution

The circular dichroism spectra were acquired using a JASCO J-1500 spectrophotometer with a 10 mm pathlength quartz at room temperature. A collection rate of 50 nm/min and a bandwidth of 1 nm were used during the measurement. SEM images were collected using a Hitachi S4800 SEM. The nanoparticles were drop-cast onto a silicon wafer and air-dried before taking SEM images.

### Microscopic circular dichroism measurement in air

For the characterization of chiral gold nanoparticles, a supercontinuum laser (SuperK Extreme, NKT Photonics) filtered by a tunable filter (SuperK VARIA, NKT Photonics) was used as the illumination source. The polarization of the illumination is controlled by the previously mentioned half-wave plate and quarter-wave plate. The reflection spectra of the chiral nanoparticles are captured by the spectrometer (LR1-B, ASEQ Instruments). The spectra were acquired 10 times with an integration time of 100 ms at 2-nm intervals using 10 sampling boxes, and the resulting data were averaged. A reference spectrum from a bare glass substrate was also collected and subtracted as background.

# References

1. Altug, H., Oh, S. H., Maier, S. A., Homola, J. Advances and applications of nanophotonic biosensors. *Nat. Nanotechnol.* **17**, 5-16 (2022).

2. Kamandi, M., et al. Enantiospecific detection of chiral nanosamples using photoinduced force. *Phys Rev Appl* **8**, (2017).

3. Hang, Y. J., Wang, A. Y., Wu, N. Q. Plasmonic silver and gold nanoparticles: Shape- and structure-modulated plasmonic functionality for point-of-caring sensing, bio-imaging and medical therapy. *Chem. Soc. Rev.* **53**, 2932-2971 (2024).

4. Yuan, L., Bourgeois, B. B., Carlin, C. C., da Jornada, F. H., Dionne, J. A. Sustainable chemistry with plasmonic photocatalysts. *Nanophotonics* **12**, 2745-2762 (2023).

5. Lin, K. T., Lin, H., Jia, B. H. Plasmonic nanostructures in photodetection, energy conversion and beyond. *Nanophotonics* **9**, 3135-3163 (2020).

6. Zhang, S. Y., et al. Metasurfaces for biomedical applications: Imaging and sensing from a nanophotonics perspective. *Nanophotonics* **10**, 259-293 (2021).

7. Im, S., Mousavi, S., Chen, Y.-S., Zhao, Y. Perspectives of chiral nanophotonics: From mechanisms to biomedical applications. *npj Nanophotonics* **1**, 46 (2024).

8. Baffou, G., Quidant, R. Thermo-plasmonics: Using metallic nanostructures as nano-sources of heat. *Laser Photonics Rev.* **7**, 171-187 (2013).

9. Govorov, A. O., Richardson, H. H. Generating heat with metal nanoparticles. *Nano Today* **2**, 30-38 (2007).

10. Chen, Y. S., et al. Silica-coated gold nanorods as photoacoustic signal nanoamplifiers. *Nano Lett.* **11**, 348-354 (2011).

11. Mateo, D., Cerrillo, J. L., Durini, S., Gascon, J. Fundamentals and applications of photo-thermal catalysis. *Chem. Soc. Rev.* **50**, 2173-2210 (2021).

12. Yang, B., Li, C. Y., Wang, Z. F., Dai, Q. Thermoplasmonics in solar energy conversion: Materials, nanostructured designs, and applications. *Adv. Mater.* **34**, 2107351 (2022).

13. Riley, R. S., Day, E. S. Gold nanoparticle-mediated photothermal therapy: Applications and opportunities for multimodal cancer treatment. *Wires. Nanomed. Nanobi.* **9**, e1449 (2017).

14. Löw, P., Kim, B., Takama, N., Bergaud, C. High-spatial-resolution surface-temperature mapping using fluorescent thermometry. *Small* **4**, 908-914 (2008).

15. Okabe, K., et al. Intracellular temperature mapping with a fluorescent polymeric thermometer and fluorescence lifetime imaging microscopy. *Nat. Commun.* **3**, 705 (2012).

16. Baffou, G., Kreuzer, M. P., Kulzer, F., Quidant, R. Temperature mapping near plasmonic nanostructures using fluorescence polarization anisotropy. *Opt. Express* **17**, 3291-3298 (2009).

17. Sapsford, K. E., Berti, L., Medintz, I. L. Materials for fluorescence resonance energy transfer analysis: Beyond traditional donor-acceptor combinations. *Angew. Chem. Int. Edit.* **45**, 4562-4588 (2006).

18. Barella, M., et al. Photothermal response of single gold nanoparticles through hyperspectral imaging anti-Stokes thermometry. *ACS Nano* **15**, 2458-2467 (2021).

19. Baffou, G., Quidant, R., de Abajo, F. J. G. Nanoscale control of optical heating in complex plasmonic systems. *ACS Nano* **4**, 709-716 (2010).

20. Xia, Q., Yin, J. Z., Guo, Z. Y., Cheng, J. X. Mid-infrared photothermal microscopy: Principle, instrumentation, and applications. *J. Phys. Chem. B* **126**, 8597-8613 (2022).

21. Mecklenburg, M., et al. Nanoscale temperature mapping in operating microelectronic devices. *Science* **347**, 629-632 (2015).

22. Dazzi, A., Prater, C. B. AFM-IR: Technology and applications in nanoscale infrared spectroscopy and chemical imaging. *Chem. Rev.* **117**, 5146-5173 (2017).

23. Jahng, J., Potma, E. O., Lee, E. S. Tip-enhanced thermal expansion force for nanoscale chemical imaging and spectroscopy in photoinduced force microscopy. *Anal Chem.* **90**, 11054-11061 (2018).

24. Almajhadi, M. A., Uddin, S. M. A., Wickramasinghe, H. K. Observation of nanoscale opto-mechanical molecular damping as the origin of spectroscopic contrast in photo induced force microscopy. *Nat. Commun.* **11**, 5691 (2020).

25. Woo, H. J., et al. Advances and challenges in dynamic photo-induced force microscopy. *Discov. Nano* **19**, 190 (2024).

26. Lu, F., Jin, M. Z., Belkin, M. A. Tip-enhanced infrared nanospectroscopy via molecular expansion force detection. *Nat. Photonics* **8**, 307-312 (2014).

27. O'Callahan, B. T., Yan, J., Menges, F., Muller, E. A., Raschke, M. B. Photoinduced tip-sample forces for chemical nanoimaging and spectroscopy. *Nano Lett.* **18**, 5499-5505 (2018).

28. Lee, H., et al. Probing temperature-induced plasmonic nonlinearity: Unveiling opto-thermal effects on light absorption and near-field enhancement. *Nano Lett.* **24**, 3598-3605 (2024).

29. Wang, H. W., Meyer, S. M., Murphy, C. J., Chen, Y. S., Zhao, Y. Visualizing ultrafast photothermal dynamics with decoupled optical force nanoscopy. *Nat. Commun.* **14**, 7267 (2023).

30. Baffou, G., et al. Deterministic temperature shaping using plasmonic nanoparticle assemblies. *Nanoscale* **6**, 8984-8989 (2014).

31. Wan, J. L., et al. $Cu^{2+}$-dominated chirality transfer from chiral molecules to concave chiral Au nanoparticles. *J. Am. Chem. Soc.* **146**, 10640-10654 (2024).

32. Ni, B., et al. Seed-mediated growth and advanced characterization of chiral gold nanorods. *Adv. Mater.* **36**, 2412473 (2024).

33. Varesi, J., Majumdar, A. Scanning joule expansion microscopy at nanometer scales. *Appl. Phys. Lett.* **72**, 37-39 (1998).